\newcommand{\CLASSINPUTtoptextmargin}{0.75in}
\newcommand{\CLASSINPUTbottomtextmargin}{1.05in}
\documentclass[conference]{IEEEtran}
\usepackage{graphicx}
\usepackage{subfigure}
\usepackage{float}
\usepackage{amsmath}
\usepackage{amsfonts,amssymb}
\usepackage{dsfont}
\usepackage{algpseudocode}
\usepackage{makecell}
\usepackage{cite}
\usepackage{multirow}
\usepackage{multicol}
\usepackage{booktabs}

\usepackage[linesnumbered, lined, ruled, noend]{algorithm2e}
\usepackage{soul}
\usepackage{tabularx}
\usepackage[table,xcdraw]{xcolor}
\usepackage{rotating}
\usepackage{afterpage}
\usepackage{pdflscape}
\usepackage{url}
\usepackage[english]{babel}
\newcommand*{\myprime}{^{\mkern 1.2mu \prime}}
\SetCommentSty{mycommfont}
\SetKwInput{KwInput}{Input}                % Set the Input
\SetKwInput{KwOutput}{Output}              % set the Output
\def\BibTeX{{\rm B\kern-.05em{\sc i\kern-.025em b}\kern-.08em
    T\kern-.1667em\lower.7ex\hbox{E}\kern-.125emX}}

\DeclareFontFamily{U}{mathx}{}
\DeclareFontShape{U}{mathx}{m}{n}{<-> mathx10}{}
\DeclareSymbolFont{mathx}{U}{mathx}{m}{n}
\DeclareMathAccent{\widehat}{0}{mathx}{"70}
\DeclareMathAccent{\widecheck}{0}{mathx}{"71}

\usepackage{fancyhdr,lipsum}
\fancypagestyle{mahmood}{%
   \fancyhf{} % clear all fields
   
   \fancyhead[C]{2025 IEEE. You can use this material personally. Reprinting or republishing this material for the purpose of advertising or promotion, creating new collective works, reselling or redistributing to servers or lists, or using any copyrighted component in other works must adhere to IEEE policy. The DOI will be supplied as soon as it becomes available.}
}%

\makeatletter
\let\ps@IEEEtitlepagestyle\ps@mahmood
\makeatother

\begin{document}

\title{Deep Learning Based Service Composition in Integrated Aerial-Terrestrial Networks}

% \title{Service Composition in Integrated Aerial/Terrestrial Networks: Deep Learning based Aerial-Base Stations Locations and Requests Predictors}

\author{
    \IEEEauthorblockN{
        Mohammad Farhoudi\textsuperscript{1}, Masoud Shokrnezhad\textsuperscript{1,2}, Somayeh Kianpisheh\textsuperscript{1}, and Tarik Taleb\textsuperscript{3}\\
    }
    \IEEEauthorblockA{
        \textsuperscript{1} \textit{Oulu University, Finland};
        \{mohammad.farhoudi, masoud.shokrnezhad, somayeh.kianpisheh\}@oulu.fi \\
        \textsuperscript{2} \textit{ICTFICIAL Oy, Espoo, Finland}; masoud.shokrnezhad@ictficial.com \\
        \textsuperscript{3} \textit{Ruhr University Bochum (RUB), Germany}; \{tarik.taleb\}@rub.de 
    }
}
\maketitle

\begin{abstract}
The explosive growth of user devices and emerging applications is driving unprecedented traffic demands, accompanied by stringent Quality of Service (QoS) requirements. Addressing these challenges necessitates innovative service orchestration methods capable of seamless integration across the edge-cloud continuum. Terrestrial network-based service orchestration methods struggle to deliver timely responses to growing traffic demands or support users with poor or lack of access to terrestrial infrastructure. Exploiting both aerial and terrestrial resources in service composition increases coverage and facilitates the use of full computing and communication potentials. This paper proposes a service placement and composition mechanism for integrated aerial-terrestrial networks over the edge-cloud continuum while considering the dynamic nature of the network. The service function placement and service orchestration are modeled in an optimization framework. Considering the dynamicity, the Aerial Base Station (ABS) trajectory might not be deterministic, and their mobility pattern might not be known as assumed knowledge. Also, service requests can traverse through access nodes due to users' mobility. By incorporating predictive algorithms, including Deep Reinforcement Learning (DRL) approaches, the proposed method predicts ABS locations and service requests. Subsequently, a heuristic isomorphic graph matching approach is proposed to enable efficient, latency-aware service orchestration. Simulation results demonstrate the efficiency of the proposed prediction and service composition schemes in terms of accuracy, cost optimization, scalability, and responsiveness, ensuring timely and reliable service delivery under diverse network conditions.
% The explosive growth of user devices and emerging applications is driving unprecedented traffic demands, accompanied by stringent quality of service requirements. Traditional terrestrial service orchestration struggles to deliver timely responses to growing traffic demands or support users with poor or lack of access to network infrastructure. Exploiting both aerial and terrestrial resources in service composition increases coverage and facilitates full computing and communication potentials. This paper proposes a dynamic service placement and composition mechanism over the integrated aerial-terrestrial edge-cloud continuum, addressing the dynamic nature of the network. Considering the dynamicity, the Aerial Base Station (ABS) trajectory might not be deterministic, and their mobility pattern might not be known as assumed knowledge. Also, service requests traverse through access nodes due to users' mobility. By incorporating deep reinforcement learning approaches, the proposed method predicts ABS locations and service requests. A Hungarian isomorphic graph matching approach is then proposed to enable latency-aware service orchestration. Simulation results demonstrate the efficiency of the proposed scheme in terms of cost optimization, scalability, and responsiveness, ensuring efficient service delivery under diverse network conditions.
\end{abstract}


\section{Introduction}
Today's landscape exhibits a considerable increase in the number of users, leading to an immense rise in traffic demand \cite{Slicing2024}. The rapid growth of emerging applications like the Industrial Internet of Things (IIoT) has further driven this demand \cite{Blockchain2023}. Resultantly, there is a need to reassess service orchestration methods to meet evolving capacity and Quality of Service (QoS) criteria \cite{ACNC}. This imperative aims to ensure the efficient allocation of resources throughout the edge-cloud continuum \cite{DistributedRA6G}. The placement of service functional block instances (or simply instances) and traffic routing emerge as challenges, particularly in light of the escalating complexity of service structures and stringent QoS demands.

\begin{figure}[t!]\centering
\vspace{-5pt}
\includegraphics[width=3.3in]{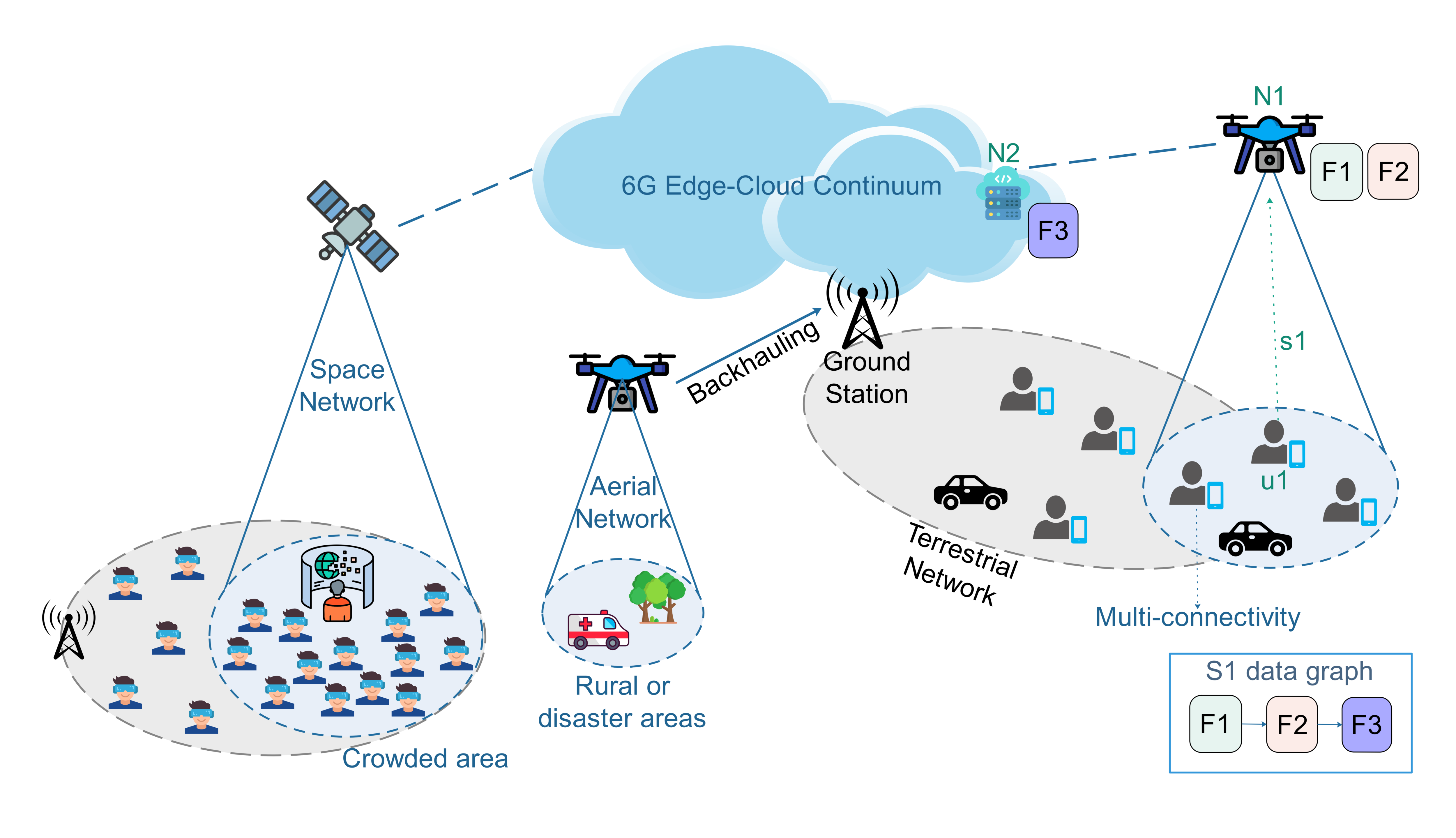}
\vspace{-0.55cm}
  \caption{A conceptual diagram illustrating the integration of aerial and terrestrial networks in the 6G edge-cloud continuum.}
    \vspace{0.1cm}
    \label{figure:architecture}
\end{figure}

Service composition solutions focusing on terrestrial networks have been studied in the literature \cite{wu2023predictive}. However, the mobile nature of users and growing traffic demand highlight the inefficiency of terrestrial network-based solutions. Unmanned Aerial Vehicles (UAVs) deployed as Aerial Base Stations (ABSs) present a viable option for leveraging edge computing resources to support users with poor or no terrestrial connectivity. This becomes particularly relevant in scenarios where ground networks encounter coverage limitations (Fig. \ref{figure:architecture}).

Some studies have been conducted on service orchestration for non-terrestrial networks. 
% Santos \textit{et al.} \cite{santos2023mobility} emphasized the deployment of UAVs in poor areas to facilitate user requests. 
Wei \textit{et al.} \cite{weiJoint2023} optimized UAV trajectory planning and service deployment in scenarios with obstacles. He \textit{et al.} \cite{he2024online} studied the joint optimization of virtualized service provisioning and UAV trajectory planning. An optimization approach for service relocation and handover in UAV networks was provided by Bekkouche \textit{et al.} \cite{bekkouche2021toward}.
% The authors of \cite{fadlullah2020hcp} provided a computing platform across the UAVs and Ground Base Stations (GBSs) for content caching services. 
% Gonzalez \textit{et al.} \cite{gonzalez2019transport} investigated how intermittent aerial communications affect the orchestration. 
Qu \textit{et al.} \cite{QuOfflineLearn} presented an offline learning-based orchestration scheme. Lastly, Wang \textit{et al.} \cite{WangPlacement2019} presented Joint Composition Assignment and Placement (Jcap), integrating service composition with Virtual Network Function (VNF) placement and assignment to enhance resource allocation efficiency.
% Their approach considered nodes’ capacity while placing services and assigning VNFs within users’ close distance, considering the maximum hop limit.

Although the above-mentioned methods address orchestration solutions, some are limited to aerial networks. Conversely, 6G network requirements encourage aerial-terrestrial resource integration to maximize network coverage and exploit the full potential of the network infrastructure. Considering the mobility of ABSs, the literature either plans the trajectory \cite{he2024online} or assumes deterministic and known mobility patterns for ABSs \cite{determinedPath2025, bekkouche2021toward}. Under multi-administrative/operator-based scenarios, ABSs are managed by several operators \cite{humann2018modeling}. Hence, UAV trajectories exhibit multiplexity or a lack of opportunity to reveal their trajectory due to conventional purposes.  
% The authors of \cite{predictUAV} discuss the non-determinism and dynamicity in UAV mobility - due to environmental factors, time delays, and other factors. 

To address the above-mentioned gaps, this paper proposes a service placement and composition mechanism for integrated aerial-terrestrial networks. To deal with the dynamic nature of ABS mobility patterns, a Deep Reinforcement Learning (DRL)-based scheme is provided to predict ABS locations. In contrast with a random stochastic behavior assumption for ABSs, exploiting the prediction results leads to a more efficient optimization of resource allocation. Also, to deal with users' mobility, an approach is followed in which requests can traverse through multiple Base Stations (BSs) and a DRL-based scheme is employed to predict requests at BSs. A heuristic method based on the Hungarian isomorphic graph matching approach is then provided to solve the near-optimal service placement and composition problem.

In the rest, Section \ref{sec:system_model} outlines the system model, and Section \ref{sec:problem} gives the optimization framework. Section \ref{sec:method} elaborates on deep learning based service placement and composition. Section \ref{sec:results} presents numerical findings, while Section \ref{sec:conclusion} offers concluding remarks and outlines future research directions.

\section{System Model} \label{sec:system_model}
% The 6G edge-cloud continuum facilitates access to services by three components: edge-cloud infrastructure, service providers, and service requests.

\textbf{Edge-Cloud Infrastructure:} The network infrastructure contains aerial and terrestrial nodes equipped with computing, storage, and networking, modeled as a network graph $\mathcal{G(\boldsymbol{\mathcal{N}}, \boldsymbol{\mathcal{L}}, \boldsymbol{\mathcal{P}})}$. Network nodes $\boldsymbol{\mathcal{N}}$ include edge-cloud, Ground Base Station (GBS), and ABS nodes. Each node has a usage cost $\overline{\mathcal{C}}_n$, remaining battery power $\widehat{\mathcal{P}}_n$, and a capacity threshold $\widehat{\mathcal{C}}_n$. The network graph contains links ($\boldsymbol{\mathcal{L}} \subset \{ l : (n, n\myprime)|n, n\myprime \in \mathcal{N} \}$), with each link's maximum bandwidth $\widehat{\mathcal{L}}_l$ and the cost of using links for packet transmission $\mathcal{\overline{L}}_l$. Users and edge-cloud nodes should exist in each other's coverage to be considered linked. To model dynamicity in the network due to ABs and users mobility, directional paths $\boldsymbol{\mathcal{P}}^t = { p: (\mathcal{H}^t_p, \mathcal{T}^t_p) | p \subset \boldsymbol{\mathcal{L}} }$ varies over time $t$. Paths are defined by their head (\( \mathcal{H}^t_p \)) and tail (\( \mathcal{T}^t_p \)) nodes, with $\mathcal{J}^t_{p,l}$ indicating whether path $p$ includes link $l$. Finally, the network architecture is tiered, with varying computing capacities in which edge nodes closer to users possess limited yet costly resources, whereas cloud nodes offer cost-effective and virtually limitless capabilities \cite{farhoudi2024}. The nodes that form a particular path $p$ are represented as $\mathcal{N}^\star_p$.

\textbf{Service Providers:} A set of service providers within the edge-cloud infrastructure offer various services, denoted by $\boldsymbol{\mathcal{S}}\!\!=\!\!\{1,2, ..., \mathcal{S}\}$, distinguished by their definitions and requirements. The requirements include composition elements - inputs, outputs, preconditions - and QoS aspects such as cost and latency. Each service $s \!\in\! \boldsymbol{S}$ has a predefined service data graph $\mathcal{G}_s$ and maintains a set of $\mathcal{C}_{s}$ atomic functional blocks, dubbed $\boldsymbol{\mathcal{F}}_s\!=\!\!\{1,2, ..., \mathcal{C}_{s}\}$. An example can be found in Fig. \ref{figure:architecture}, where user $u_1$ sends $s_1$ request with the data graph $\mathcal{G}_{s1}$, which has its functions deployed on $N_1$ and $N_2$. The services have an overall time requirement of $\mathcal{W}_s$ to orchestrate the organizational structure. Besides, each function $f \!\in\! \boldsymbol{\mathcal{F}}_s$ is implemented by instantiating an instance from $\boldsymbol{\mathcal{I}}_f\!=\!\!\{1,2,...,\mathcal{I}_f\}$, each of which capable of managing multiple requests at the cost of use $\overline{\mathcal{I}}_{f,i}$ but constrained by a capacity threshold $\widehat{\mathcal{I}}_{f,i}$.

\textbf{Service Requests:}
A group of users initiates service requests represented by $\boldsymbol{\mathcal{R}}\!\!=\!\!\{1,2, ..., \mathcal{R}\}$. Each request enters the system at time $\mathcal{T}_e$, demands a service denoted as $\mathcal{S}_r$, and has a request lifetime $\mathcal{W}_s$. Users exhibit dynamic behavior, with their locations varying over time as tracked by \( \mathcal{L}^{t}_{r} \) and their Point of Attachment (PoA)--the user's network gateway--identified at each time slot \( \mathcal{E}^{t}_{r} \). Upon request initiation, the selection of the most appropriate atomic service function instances for the request $r$ becomes paramount. This selection process takes into account various service requirements, including the minimum capacity for each function $\widecheck{\mathcal{I}}^{t}_{r,f}$, minimum network bandwidth $\widecheck{\mathcal{L}}^{t}_{r}$, maximum acceptable End-to-End (E2E) latency $\widecheck{\mathcal{D}}^{t}_{r}$, traffic burst $\widecheck{\mathcal{B}}^{t}_{r}$, and maximum packet size for each function $\widecheck{\mathcal{Z}}^{t}_{r,f}$. It also considers $\widecheck{\mathcal{Y}}_{r}$, which signifies the upper limit of tolerable overall E2E latency for requests throughout $\mathcal{W}_s$ time slots commonly referred to as the Service-Level Agreement (SLA) requirement \cite{globecom2023}.

\section{Optimization Framework} \label{sec:problem}
The service placement and composition problem is a Mixed-Integer Nonlinear Programming (MINLP) optimization with the formulation defined in (\ref{certainest}). The formulation includes function instance placement for deploying them on nodes, path selection, functions assignment, and constraints associated with resources and service requirements. The primary goal (OF) is to maximize service request acceptance while reducing costs within a given period $\boldsymbol{\mathcal{T}}$. The binary variables $\Ddot{\mathcal{A}}^{t}_{r,f,i}$ and $\Ddot{\mathcal{E}}^{t}_{f,i,n}$ signify the selection of instance $i$ of function $f$ for request $r$, and the selection of the hosting node $n$ for instance $i$, respectively. Scaling factor $\Psi$ adjusts the relative impact of criteria for service request acceptance and total computational, request forwarding, and deployment costs.

\vspace{-6pt}
\footnotesize
\begin{align}\label{certainest}
    & \mathrm{ max } \quad \text{ OF } \quad \mathrm{ s.t } \quad \text{C1 - C11.} \tag{1} \\
    & \!\!\! \sum_{ \boldsymbol{\mathcal{T}},\boldsymbol{\mathcal{R}}, \boldsymbol{\mathcal{F}}_{s_r}, \boldsymbol{\mathcal{I}}_f } \!\!\!\!\!\!\!\!\!\!\!\! \Ddot{\mathcal{A}}^{t}_{r,f,i} - \! \Psi \left( \! \sum_{\boldsymbol{\mathcal{T}}, \boldsymbol{\mathcal{F}}_{s}, \boldsymbol{\mathcal{I}}_f, \boldsymbol{\mathcal{N}}} \!\!\!\!\!\!\!\!\!   \Ddot{\mathcal{E}}^{t}_{f,i,n} \overline{\mathcal{C}}_n + \!\!\!\!\!\!\!\!\!\!\!\! \sum_{\boldsymbol{\mathcal{T}}, \boldsymbol{\mathcal{R}}, \boldsymbol{\mathcal{F}}_{s_r}, \boldsymbol{\mathcal{I}}_f}\!\!\!\!\!\!\!\!\!\!\!\! \Ddot{\mathcal{A}}^{t}_{r,f,i} \overline{\mathcal{I}}_{f,i} \!+ \!\! \sum_{\boldsymbol{\mathcal{T}}, \boldsymbol{\mathcal{R}}} \!\! \Ddot{\mathcal{L}}^{t}_{r} \! \right) \tag{OF}
\end{align}
\normalsize

\textbf{Resource Allocation:}
Allocating resources involves deploying function instances on network nodes, assigning requests to deployed instances, and routing packets within a capacity-constrained environment. Constraint (C1) assures that each request belongs to a unique instance of a function. Constraint (C2) ensures that each function instance selected by a request should be placed on an available infrastructure node for the duration of the requested service $\boldsymbol{\mathcal{T}}_r = [\mathcal{T}_e,\mathcal{T}_e + \mathcal{W}_s]$.

\vspace{-5pt}
\footnotesize
\begin{align}\label{resource_allocation}
    & \sum_{ \boldsymbol{\mathcal{I}}_f } \Ddot{\mathcal{A}}^{t}_{r,f,i} \leq 1 \qquad \qquad \qquad \qquad \qquad \forall r,f,t \in \boldsymbol{\mathcal{R}}, \boldsymbol{\mathcal{F}}_{s_r}, \boldsymbol{\mathcal{T}}_{r} , \tag{C1}
    \\
    & \sum_{\boldsymbol{\mathcal{N}}} \Ddot{\mathcal{E}}^{t}_{f,i,n} > ( \sum_{\boldsymbol{\mathcal{R}}} \Ddot{\mathcal{A}}^{t}_{r,f,i} ) / \mathcal{R} \qquad \quad \forall f, i, t \in \boldsymbol{\mathcal{F}}_s, \boldsymbol{\mathcal{I}}_f, \boldsymbol{\mathcal{T}}. \tag{C2}
\end{align}
% \vspace{-3mm}
\normalsize

Maintaining stability while handling user demands and network conditions requires capacity constraints. Given the finite capacity, storage, and computational resources, the total number of requests assigned to each service instance (C3) and deployed on each node (C4) should not surpass capacity limits.
% As ABSs have limited battery power, a power constraint that restricts the energy expended is needed. Using constraint (C5), energy consumption is limited during instance deployment and packet transmission, which is proportional to cost.

\vspace{-5pt}
\footnotesize
\begin{align}\label{resource_allocation2}
    &\sum_{\boldsymbol{\mathcal{R}}} \Ddot{\mathcal{A}}^{t}_{r,f,i} \widecheck{\mathcal{I}}^{t}_{r,f} \leq \widehat{\mathcal{I}}_{f,i} \qquad \qquad \qquad \qquad \quad \forall f, i, t \in \boldsymbol{\mathcal{F}}_s, \boldsymbol{\mathcal{I}}_f, \boldsymbol{\mathcal{T}}, \tag{C3}
    \\
    &\sum_{\boldsymbol{\mathcal{R}}, \boldsymbol{\mathcal{F}}_{s_r}, \boldsymbol{\mathcal{I}}_f }\Ddot{\mathcal{E}}^{t}_{f,i,n} \Ddot{\mathcal{A}}^{t}_{r,f,i} \widecheck{\mathcal{I}}^{t}_{r,f} \leq \widehat{\mathcal{C}}_n \qquad \qquad \qquad \quad \forall n, t \in \boldsymbol{\mathcal{N}}, \boldsymbol{\mathcal{T}}, \tag{C4} 
    % \\
    % & \sum_{\boldsymbol{\mathcal{T}}, \boldsymbol{\mathcal{F}}_s, \boldsymbol{\mathcal{I}}_f } \!\!\! \Ddot{\mathcal{E}}^{t}_{f,i,n} \overline{\mathcal{C}_n}  + \!\!\!\!\!\!\!\!\!\! \sum_{\boldsymbol{\mathcal{T}}, \boldsymbol{\mathcal{R}}, \boldsymbol{\mathcal{L}}, \boldsymbol{\mathcal{F}_{s_r}},\boldsymbol{\mathcal{P}}^t | \mathcal{H}^t_p=n} \!\!\!\!\!\!\!\!\!\!\!\!\!\!\!\!\! \overline{\mathcal{L}_l}\mathcal{J}^t_{p,l} \mathcal{R}^{t}_{r,f,p} \leq \widehat{\mathcal{P}}_n \quad \forall n \in \boldsymbol{\mathcal{N}}. \tag{C5}
\end{align}
% \vspace{-2mm}
\normalsize

Establishing feasible E2E routes for each request is necessary to facilitate the transmission of inquiry traffic from a user to its designated instances and the return of the response. ABS movements result in a variation of paths \( \boldsymbol{\mathcal{P}}^t \) during different time slots. To achieve round-trip path selection, a distinct inquiry path is chosen for each request. It originates at its network's entry node (PoA) and culminates at the specified first function deployed node. This path traverses through the other functions in an order adhering to $\mathcal{G}_s$ and returns to the PoA (C5). Binary variable $\overrightarrow{\mathcal{R}}^{t}_{r,f,p}$ indicates the assignment of path $p$ for traffic steering to the function $f$ of request $r$. Capacity limitation (C6) regulates the number of requests assigned to each link at any given time, ensuring optimal path allocation and further optimizing link allocation efficiency. Using (C7), we calculate the total costs associated with (OF).

\vspace{-5pt}
\footnotesize
\begin{align}\label{resource_allocation3}
    & \sum_{\boldsymbol{\mathcal{P}}^t | \mathcal{H}^t_p=\mathcal{E}^{\mathcal{T}_e}_{u_r} \; \& \; \mathcal{T}^t_p = \mathcal{E}^{\mathcal{T}_r}_{u_r} \; \& \; \forall n,f \in \mathcal{N}^\star_p,\mathcal{C}_{s_r}} \!\!\!\!\!\!\!\!\!\!\!\!\!\!\!\!\!\!\!\!\!\!\!\!\!\!\!\!\!\!\!\!\!\!\!\!\!\!\!\!\!\!\! \overrightarrow{\mathcal{R}}^{t}_{r,f,p}(\Ddot{\mathcal{E}}^t_{i,f,n} == 1) = 1 \quad \forall r, t \!\! \in \boldsymbol{\mathcal{R}}, \boldsymbol{\mathcal{T}}_r, \tag{C5} 
    \\
    &\sum_{\boldsymbol{\mathcal{R}}} \widecheck{\mathcal{L}}^{t}_{r} \sum_{\boldsymbol{\mathcal{F}_{s_r}},\boldsymbol{\mathcal{P}}^t} \mathcal{J}^t_{p,l} \overrightarrow{\mathcal{R}}^{t}_{r,f,p} \leq \widehat{\mathcal{L}}_{l} \qquad \qquad \qquad \qquad \forall l, t \in \boldsymbol{\mathcal{L}}, \boldsymbol{\mathcal{T}}, \tag{C6}
    \\
    &\Ddot{\mathcal{L}}^{t}_{r} = \sum_{\boldsymbol{\mathcal{L}}} \overline{\mathcal{L}}_l\sum_{\boldsymbol{\mathcal{F}_{s_r}}, \boldsymbol{\mathcal{P}}^t} \mathcal{J}^t_{p,l} \overrightarrow{\mathcal{R}}^{t}_{r,f,p} \qquad \qquad \qquad \qquad \forall r, t \in \boldsymbol{\mathcal{R}}, \boldsymbol{\mathcal{T}}. \tag{C7}
\end{align}
% \vspace{-3mm}
\normalsize

\textbf{QoS Requirements:}
In each time slot, the maximum acceptable latency should be maintained to ensure compliance with E2E latency thresholds (C10). Also, the cumulative latencies experienced by requests from each user across all time slots from different atomic service instances should not exceed the request SLA (C11). Continuous variables $\mathcal{D}^{t}_{r,l}$ and $\mathcal{D}^{t}_{r}$ quantify the latency of link $l$ and each request's E2E latency respectively, including both network and computing latencies for $r$ \cite{shokrnezhad2022}. These constraints collectively safeguard timely and reliable service delivery, maintaining a high-performance standard in line with users' stringent requirements.

\vspace{-5pt}
\footnotesize
\begin{align}\label{qos_req}
& \mathcal{D}^{t}_{r,l} = ( \sum_{\boldsymbol{\mathcal{R}} | r\myprime \neq r} \widecheck{\mathcal{B}}^{t}_{r'} + \sum_{\boldsymbol{\mathcal{F}}_{s_r}}\widecheck{\mathcal{Z}}^{t}_{r',f} ) / \widehat{\mathcal{L}}_{l} \qquad \qquad \forall r, l, t \in \boldsymbol{\mathcal{R}}, \boldsymbol{\mathcal{L}}, \boldsymbol{\mathcal{T}}, \tag{C8} 
    \\
    & \mathcal{D}^{t}_{r} = \!\!\!\!\! \sum_{\boldsymbol{\mathcal{F}_{s_r}},\boldsymbol{ \mathcal{P}}^t, \boldsymbol{\mathcal{L}}} \mathcal{J}^t_{p,l} \mathcal{D}^{t}_{r,l}  \overrightarrow{\mathcal{R}}^{t}_{r,f,p} + \sum_{ \boldsymbol{\mathcal{F}}_{s_r} }\widecheck{\mathcal{Z}}^{t}_{r,f}/\widecheck{\mathcal{I}}^{t}_{r,f} \quad \forall r, t \in \boldsymbol{\mathcal{R}}, \boldsymbol{\mathcal{T}}, \tag{C9}
    \\
    & \mathcal{D}^{t}_{r} \leq \widecheck{\mathcal{D}}^{t}_{r} \qquad \qquad \qquad \qquad \qquad \qquad \qquad \qquad \quad \forall r, t \in \boldsymbol{\mathcal{R}}, \boldsymbol{\mathcal{T}}, \tag{C10}
    \\
    & \sum_{ \boldsymbol{\mathcal{T}}_{r}} \mathcal{D}^{t}_{r} \leq \widecheck{\mathcal{Y}}_{r} \qquad \qquad \qquad \qquad \qquad \qquad \quad \qquad \qquad \forall r \in \boldsymbol{\mathcal{R}}. \tag{C11}
\end{align}
\normalsize

% \section{Deep Learning based Service Placement and Composition} \label{sec:method}
\section{Proposed Method} \label{sec:method}
The problem of service function placement and composition is reduced to the multidimensional knapsack problem and shown to be NP-hard \cite{faticanti_cutting_2018}. The total number of possible permutations of placements in (\ref{certainest}) is of order $\mathcal{R}!\mathcal{T}!\mathcal{N}\mathcal{S}\mathcal{C}_s\mathcal{P}$ which also illustrates the complexity of the problem. To deal with the dynamic nature of the network due to the non-determinism of ABS locations, as well as the dynamicity in requests' arrival due to users' mobility, a proactive service composition scheme called a predIction based huNgariaN isOmorphic serVice orchestrATION (INNOVATION) is provided. This method addresses imperfect knowledge constraints, overcomes problem complexity, and ensures high-quality service delivery. We adopt Dueling Double Deep Q-Learning (D3QL), since it overcomes over-optimistic and unstable approximations of Q-values by exploiting two separate Q-networks.

\begin{figure}[t!]\centering
\vspace{-13pt}
\includegraphics[width=3in]{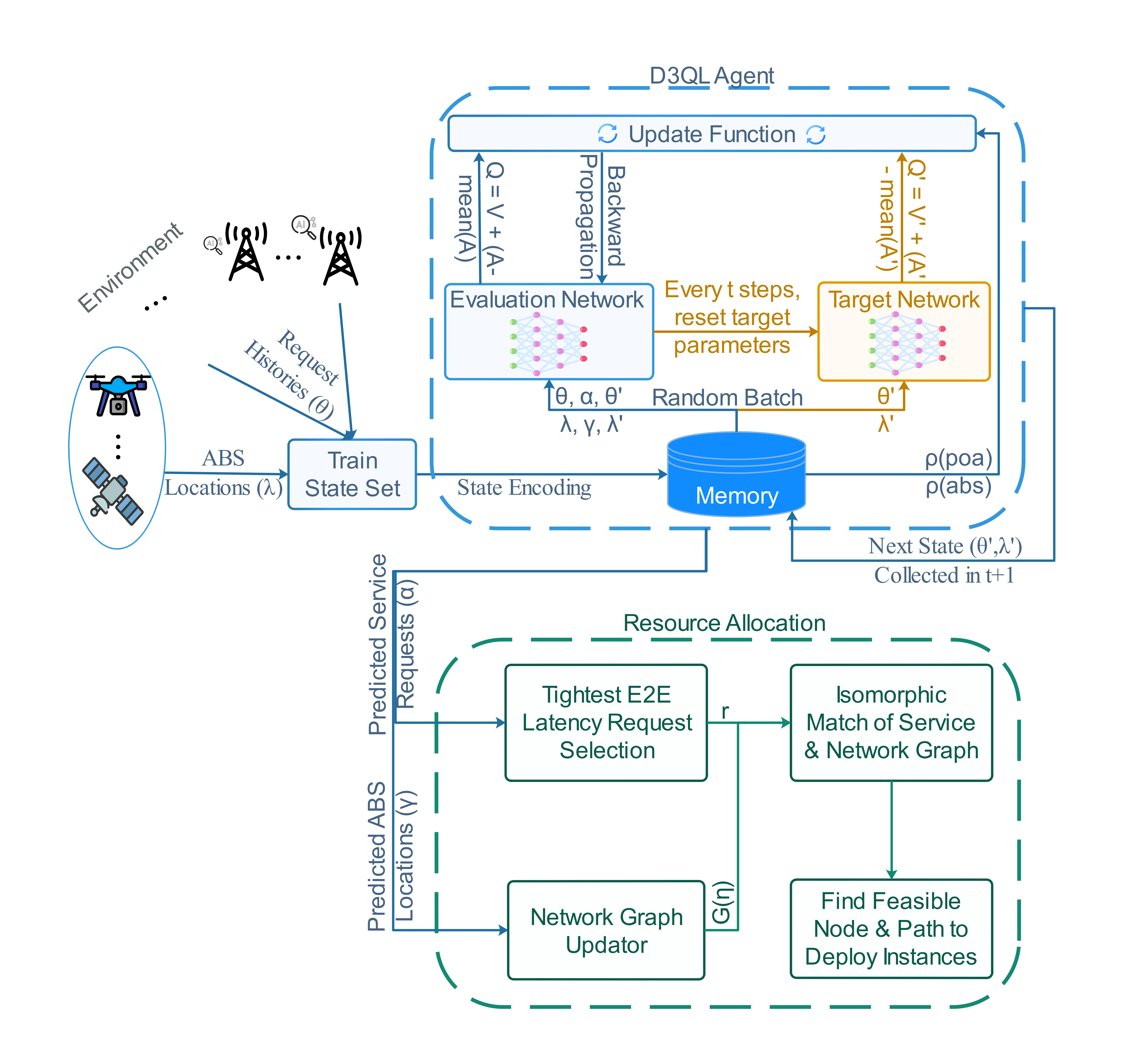}
\vspace{-18pt}
  \caption{INNOVATION learning algorithm receives environment responses, stores them, and updates the evaluation network.}
    \vspace{0.1cm}
    \label{figure:method}
\end{figure}

% Recently, Reinforcement Learning (RL) has been shown to be promising in pattern recognition in time-series problems \cite{gao2022reinforcement, kuremoto2019training}. To solve the optimization problem in Section \ref{sec:problem}, the ABS locations and user request arrivals are required to be given in a time-series manner. This demonstrates the potential of RL employment for value prediction. While RL methods can address various problems, they often struggle with efficiently handling high-dimensional state and action spaces due to the need for feature representation. To deal with this issue, we adopt DRL, which employs neural networks to approximate value functions more effectively. Particularly, we employ the advanced version of Dueling Double Deep Q-Learning (D3QL) since it overcomes over-optimistic and unstable approximation of Q-values by exploiting two separate Q-networks to select and evaluate actions \cite{hasselt_deep_2016}.

\textbf{ABS Locations Prediction:} 
The edge-cloud environment is divided into zones, with ABSs connected to the core network. Each ABS employs a D3QL agent to estimate ABS zone distribution (Algorithm \ref{Prediction}).
% For example, consider a network with four zones and an ABS. For an interval of three time slots, when the ABS moves through zones 1, 3, and 4, the sequence of states $\{1, 0, 0, 0 \}$, $\{0, 0, 1, 0 \}$, $\{0, 0, 0, 1 \}$ are given as input to the D3QL's Neural Networks (NNs), each imposing an update to the NN weights. Further enhancement of learning is achieved by storing observed transitions in a memory bank and facilitating neural network updates from this repository by random sampling.
The Markov Decision Process (MDP) state $\lambda$ represents ABS location histories, while the MDP action $\gamma$ indicates the anticipated zone for the next time slot. Each neuron in the output layer of the D3QL Neural Networks (NNs) represents the probability of being in a specific zone, predicted proportionately to the Q-values using an $\epsilon$-greedy strategy. The value of $\epsilon$ is high for exploration in early iterations and decreases linearly by $\epsilon'$ for exploitation. To motivate high accuracy gains, the MDP \textit{reward} is defined based on prediction performance. The reward is 1 when the forecast and actual zones are the same, and -1 when they are different. The D3QL agent uses two Q-networks for action selection and evaluation (Fig. \ref{figure:method}).
% The network is evolved within episodes and the NN weights are updated using the Adam optimizer \cite{AdamOptimizer}, with a periodic update of target network weights using evaluation network weights.   
After predicting ABS locations, the network graph and paths $\boldsymbol{\mathcal{P}}^t$ are updated by removing outdated links to the core network and establishing new links at anticipated locations.

% \textbf{Network Topology Reconfiguration:} After predicting ABS locations, the network graph and paths $\boldsymbol{\mathcal{P}}^t$ should be updated to reflect the revised configuration. This involves removing the links between the ABSs and the core network that are anticipated to be outdated and establishing updated links at their anticipated locations. 
% \blue{The continuous topology reconfiguration ensures that the objective function is optimized based on the learned mobility patterns of the ABSs.}

\textbf{Service Requests Prediction:} As users move, service request demands vary time-wise. Users under its coverage, which varies with their movement, determine the demand for a specific service from a BS at a particular time slot.
% As an example, a specific service that is required at a time slot may no longer be needed at another time slot if users within that BS no longer require it.
For efficient service function placement, the prediction of service request arrivals is required. At each network node that operates as a PoA, a D3QL agent is employed to predict the distribution of the request's arrival, i.e., the probability that a request $r$ is being requested at the next time slot (Algorithm \ref{Prediction}). Each agent considers state $\theta$ as the vector of arrived requests to the PoA during the last $m$ time slot. 
% With $m=1$, five existing composed services, and request arrivals of services two and four, the state is $\{0, 1, 0, 1, 0 \}$. 
The agent provides the action $\alpha$ returning a list of $z$ requests with the highest likelihood.
% Considering $z=2$ and DRL's Q-values $\{0.1, 0.9, 0.2, 0.6, 0.3\}$, the action is $\{2,4\}$, showcasing the anticipation of requests from services two and four for the next time slot. 
Finally, the reward $\rho_{poa}^t$ is assigned based on arrival requests' prediction accuracy. Comparing the predicted and the actual arrived requests, rewards 1 and 0 are issued for correct and incorrect predictions, respectively.

\begin{algorithm}[!b]
\footnotesize
\caption{DRL method for predictions}
\label{Prediction}
\KwInput{$\tau, \epsilon$, $\epsilon'$, $\widetilde{\epsilon}$, $\theta_0 \gets \{\}$, $\lambda_0 \gets \{\}$, $mem_r \gets \{\}$, $mem_a \gets \{\}$}
\KwResult{$\alpha_{\tau+1}, \gamma_{\tau+1}$}
update states ($\theta_{\tau} , \lambda_{\tau}$) $\gets$ PoA's requests \& ABS' locations \\
    \If{$\tau < m$}
    {
        $\alpha_{\tau+1}, \gamma_{\tau+1} \gets $ random $z$ services, random locations

    }
    \Else
    {
        $\zeta \gets$ generate a random number from $[0:1]$ \\
        \If{$\zeta > \epsilon$}
        {
            $\alpha_{\tau+1}, \gamma_{\tau+1} \gets$ select top Q-values
        }
        \Else
        {
            $\alpha_{\tau+1}, \gamma_{\tau+1} \gets$ select random values \\
        }
        calculate rewards ($\rho_{poa}^{\tau}$, $\rho_{abs}^{\tau}$) \\
        $mem_r \gets mem_r \cup \{(\theta_{\tau-1}, \alpha_{\tau-1}, \rho_{poa}^{\tau},\theta_{\tau})\}$ \\
        $mem_a \gets mem_a \cup \{(\lambda_{\tau-1}, \gamma_{\tau-1}, \rho_{abs}^{\tau},\lambda_{\tau})\}$ \\
        choose a sample form $mem_r$, $mem_a$ and train \\
        \If{$\epsilon > \widetilde{\epsilon}$}
        {
            $\epsilon \gets \epsilon - \epsilon'$
        }
    }

\end{algorithm}
\normalsize

\textbf{Service Placement and Resource Allocation:}
This phase focuses on determining optimal service function placements, assignments, and traffic steering for predicted requests across the edge-cloud continuum. A heuristic method is provided to allocate resources based on anticipated requests and the expected network environment. The service graph consists of nodes denoting atomic functions and edges representing data flows or dependencies between these functions. Similarly, the network graph contains network nodes such as ABSs, GBSs, and edge-cloud nodes interconnected by links with latency and capacity constraints. Isomorphic graph matching aligns the service graph with the network graph by identifying a feasible correspondence between their nodes and edges. This mapping ensures service functional relationships and dependencies are preserved while satisfying resource constraints.
% For instance, if a service graph contains three functions $f_1$, $f_2$, and $f_3$, with dependencies between $f_1$ and $f_2$, and between $f_2$ and $f_3$, the isomorphic mapping ensures that these dependencies are respected when deployed across the network. In a network graph with nodes $n_1$, $n_2$, $n_3$, and $n_4$, one feasible deployment might involve placing $f_1$ on $n_1$, $f_2$ on $n_2$, and $f_3$ on $n_3$, provided connectivity, capacity, and latency constraints are met.

The proposed method employs a Hungarian heuristic algorithm to prioritize requests with the most stringent latency needs. This prioritization ensures that time-sensitive requests are addressed first, reducing QoS violations. For each request, the algorithm evaluates candidate nodes $\overline{\mathcal{F}}(\boldsymbol{\eta})$ based on computational capacity, E2E latency, and connectivity. The total deployment cost is the sum of the node (computational) and path (communication) costs. The optimal node-path combination with the lowest cost that meets all constraints is selected for deployment (steps 7–9).
% For example, consider a scenario where the network graph consists of three nodes with computational capacities of $\{1, 13, 10\}$ and associated costs of $\{5, 13, 10\}$. A service request requires deploying a function with a resource demand of 5 units to reach the nodes. In this case, the algorithm excludes the first node due to insufficient capacity and selects the third node as the optimal placement, as it satisfies both capacity and latency requirements while minimizing cost.
After selecting optimal nodes in $\overline{\mathcal{F}}$, the method deploys function instances and establishes communication paths (steps 10–17). Existing instances are reused to improve resource utilization, and if no feasible node or path is available, the algorithm searches for alternatives. The network graph is dynamically updated based on predicted ABS locations. The iterative nature of the algorithm ensures that all service requests are processed, with priority given to the most critical ones. If a request cannot be supported due to resource limitations, it is flagged as unsupported. 
A global variable $\sigma$ tracks instance placements, enabling efficient resource reuse, reducing redundancy, and streamlining the solution space. 
% The algorithm iterates until all requests are addressed.

% Following the selection of optimal nodes in $\overline{\mathcal{F}}$, the proposed method deploys the service function instances and establishes communication paths to enable seamless data flow between interconnected functions (steps 10–17). If a suitable function instance is already deployed in the network, it is reused to avoid redundancy and improve resource utilization. Also, if a feasible node or path to the node is unavailable, the algorithm iterates over other nodes within $\overline{\mathcal{F}}$, seeking the most suitable alternative. The algorithm also dynamically updates the network graph based on the predicted locations of ABSs. The iterative nature of the algorithm ensures that all service requests are processed, with priority given to the most critical ones. If a request cannot be supported due to resource limitations, it is flagged as unsupported. Additionally, the algorithm tracks the placement of deployed instances using a global variable $\sigma$, which facilitates efficient resource reuse in subsequent iterations. This enables the utilization of shared instances when sufficient capacity is available, reducing redundancy and streamlining the solution space. The algorithm iterates until all requests are addressed.

\begin{algorithm}[!b]
\footnotesize
\caption{Hungarian isomorphic service placement}
\label{INNOVATION}

\KwInput{$\mathcal{T}$, $\alpha_0 \gets \{\}$, $\gamma_0 \gets \{\}$, $\sigma \gets \{\}$}

\For{each $\tau$ in $[1:\mathcal{T}]$}
{
    $\alpha_{\tau+1}, \gamma_{\tau+1}$ $\gets$ Predict Requests and ABS locs (Algorithm \ref{Prediction})
    
    update $\mathcal{G(\boldsymbol{\mathcal{N}}, \boldsymbol{\mathcal{L}}, \boldsymbol{\mathcal{P}}^\tau)}$ using ABS locations ($\gamma_{\tau+1}$) \\
    
    $(\boldsymbol{\mathcal{R}}$, PoAs) $ \gets $ collect $\alpha_{\tau+1}$ of all PoAs \& convert to table \\
    
    \While{$\boldsymbol{\mathcal{R}}$ is not empty} {
        $r \gets$ the tightest E2E latency required request \\

        % $min cost \gets \infty$ \\

        \For{function $f$ \& isomorphic match $\mathcal{G(\boldsymbol{\mathcal{\eta}}})$ \& $\mathcal{G}_s$ } 
        {
            \If{ $f$ in $\sigma$ \& feasible $n$,$i$ (node \& instance) in $\sigma$}{
                $\phi_1 \gets \overline{\mathcal{C}}_{n} + \overline{\mathcal{I}}_{f,i}$  \\
            }
            
            \For{set of nodes $n_f \in \overline{\mathcal{F}}(\boldsymbol{\eta})$}
            {
                $\boldsymbol{\mathcal{P}} \gets$ set of paths between $n_f$, $n$ \\
                % and PoA with first and last nodes \\
                $p_m \gets$ lowest latency $p \in \boldsymbol{\mathcal{P}}$ \\
                $\phi_2 \gets \overline{\mathcal{C}}_{n_f} + \overline{\mathcal{I}}_{f,i} + \sum_{p_m} \Ddot{\mathcal{L}}^{\tau}_{r} $  \\
                $\mathcal{D}_{p} \gets$ $p_m$'s links latency \\
            }
            
            \If{$\phi_1 + \phi_2 < min cost$ \& $\mathcal{D}_{p} \leq \widecheck{\mathcal{D}}^{\tau}_{r} $}{
                $min cost = \phi_1 + \phi_2$ \\
                $\chi$ = \{$f$\}, \{$n$ + $n_f$\}, \{$i$\}, \{$p_m$\}
            }
        }

        % \If{$min cost$ is not $\infty$}{
            $\Ddot{\mathcal{A}}^{\tau}_{r,\chi_f,\chi_i} \gets 1, 
            \Ddot{\mathcal{E}}^{\tau}_{\chi_f,\chi_i,\chi_n},
            \overrightarrow{\mathcal{R}}^{\tau}_{r,\overrightarrow{\chi_p}} \gets 1$ \\
            remove $r$ from $\boldsymbol{\mathcal{R}}$
            
            \For{each function $f$ in $\boldsymbol{\mathcal{F}}_{s_r}$ of $\mathcal{G}_{s_r}$}{
                % \If{$\widehat{\mathcal{C}}_n > 0$}{
                    $\sigma \gets \sigma +$ \{$f$ : $n$, $i$ (selected node and instance) \} 
                % }
            }
        
    }
}
\end{algorithm}
\normalsize

% \begin{table}[b!]
% \caption{Simulation Parameters}
% \vspace{-10pt}
% \label{table:simulation_paramter}
% \begin{center}
% \begin{tabular}{|c|c|}
% \hline
% \textbf{Parameter} & \textbf{Value} \\
% \hline

% link numbers ($\mathcal{L}$) &
% $\sim \mathcal{U}\{3\mathcal{N}, 5\mathcal{N}\}$ \\
% service numbers ($\mathcal{S}$) & $20$ \\
% ABS zone numbers & $25$ \\
% function instance numbers ($\mathcal{I}_f$) & $\sim \mathcal{U}\{3, 6\}$ \\
% agent history length ($m$) & $100$ \\
% link costs ($\mathcal{\overline{L}}_l$) & $\sim \mathcal{U}\{10, 20\}$ \\
% node costs ($\mathcal{\overline{C}}_n$) & $\sim \mathcal{U}\{50, 70\}$ \\
% instance costs ($\mathcal{\overline{I}}_{f,i}$) & $\sim \mathcal{U}\{25, 40\}$ \\
% link capacities ($\widehat{\mathcal{L}}_l$) & $\sim \mathcal{U}\{100, 150\}$ \\
% node capacities ($\widehat{\mathcal{C}}_n$) & $\sim \mathcal{U}\{50, 100\}$ \\
% instance capacities ($\widehat{\mathcal{I}}_{f,i}$) & $\sim \mathcal{U}\{20, 50\}$ \\
% \multirow{2}{*}{user \& ABS mobility pattern} & based on SUMO mobility \\
% & trace generation simulator \cite{SUMO2018}  \\
% \hline
% \end{tabular}
% \end{center}
% \vspace{-0.4cm}
% \end{table}

\begin{figure*}[t!]\centering
% \vspace{3pt}
\includegraphics[width=6.8in]{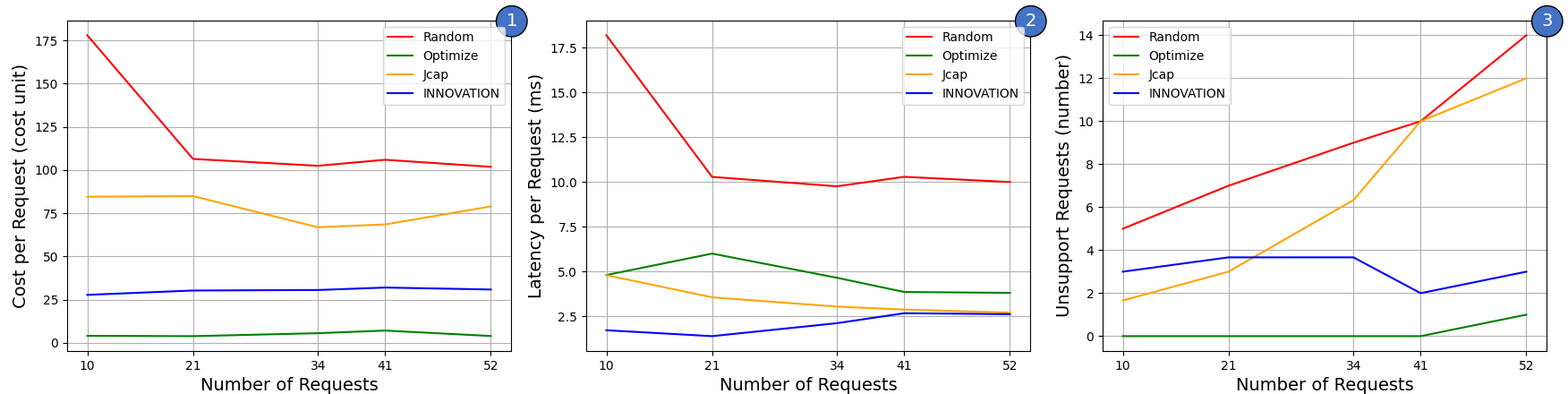}
\vspace{-8pt}
  \caption{(1) cost per request, (2) E2E latency incurred per request, and (3) unsupported request numbers, while the number of requests is set to expand.}
    \label{figure:simulation_results_requests}
    \vspace{-12pt}
\end{figure*}

\vspace{-4pt}
\section{Simulation Results}\label{sec:results}
This section evaluates the proposed method's performance. We compare our proposed method with the following methods: 1) the optimal solution that possesses omniscient knowledge of ABS locations and actual requests in advance; 2) a random strategy in which function instances and nodes are placed randomly and assigned to service requests randomly; 3) Jcap method \cite{WangPlacement2019} that composes a feasible composition by considering nodes' capacity while placing services and assigning VNFs within users' close distance.
The request numbers are strategically increased to assess the scalability of the proposed method. By subjecting our method to a diverse range of request volumes, from moderate to intense, we aim to examine its ability to efficiently allocate resources and meet SLA requirements across fluctuating demand levels. 
% \sout{It aims to reflect real-world scenarios like the Metaverse multiverses \cite{haoyu2023xr}, where network traffic can exhibit considerable fluctuations.} 
The outcomes of comparative methods are illustrated in Fig. \ref{figure:simulation_results_requests}. 

Fig. \ref{figure:simulation_results_requests}.1 depicts the deployment and resource allocation costs per request. The optimized method, knowing mobility patterns, achieves the lowest cost through cost optimization. Jcap favors low-cost cloud nodes, achieving lower costs than other methods. INNOVATION closely matches the optimized method by accurately estimating ABS locations and service request arrivals, as well as a cost-driven approach to resource allocation. As the number of requests increases, the cost of the proposed INNOVATION method remains relatively stable.

Fig. \ref{figure:simulation_results_requests}.2 illustrates INNOVATION’s low latency for accepted requests, achieved by the latency-aware heuristic algorithm. It improves latency up to 20\% over the optimized method. Jcap, despite its multi-hop access to functions, overlooks latency in resource allocation, leading to longer paths and higher latency. High latency is experienced by the optimized method since it focuses on cost optimization while ensuring deadline satisfaction. As expected, random placement and assignment of functions have the highest latency, as arbitrary resource allocation and inefficient routing result in prolonged data transmission times. Overall, INNOVATION maintains SLA-compliant latency, with only slight increases as requests grow.

Fig. \ref{figure:simulation_results_requests}.3 shows the number of unsupported requests, which serves as a metric for evaluating service continuity. The optimized method supports the most number of requests due to its omniscient approach. INNOVATION performs competitively, with fewer than 4 unsupported requests, leveraging proactive resource allocation, low latency achievement, and non-terrestrial resource exploitation. Jcap's failure to account for actual request dynamics renders it inadequate at handling growing user demands as the number of requests expands. The random method fares the worst, with a large proportion of unsupported requests due to its arbitrary allocation strategy, leading to frequent resource shortages and unmet requests.

% \begin{figure*}[t!]\centering
% % \vspace{-0.1cm}
% \includegraphics[width=7in]{Figures/combine_nodes.png}
% % \vspace{-0.3cm}
%   \caption{(1) Cost, (2) E2E latency, and (3) the number of unsupported requests are compared between the INNOVATION method and the optimized, random allocation, and Jcap methods \cite{WangPlacement2019} as network size increases.}
%     % \vspace{-0.6cm}
%     \label{figure:simulation_results_nodes}
% \end{figure*}

INNOVATION achieves an average total cost of 90\% of the optimal method. It excels at near-optimal service placement, ensuring low cost and latency regardless of request volume.
% However, variations in capacity and demand may affect cost and latency due to fluctuations in node, instance, and link capacities, and shifts in the minimum required requests' capacity and bandwidth. 
% \sout{Unsupported requests' numbers may marginally grow with the increasing requests, as INNOVATION's prediction phase anticipates a certain number of requests in the next time slot ($m$).} 
Unlike the optimized method, which prioritizes cost reduction through comprehensive scenario examination, the proposed method reduce solution space and excels at providing timely responses, on average 86\% faster than optimal. Overall, INNOVATION emerges as an efficient and scalable solution, marked by low latency, cost-effectiveness, and timely responsiveness.

% \vspace{-2pt}
\section{Conclusion}\label{sec:conclusion}
% \vspace{-2pt}
This study proposed a service placement and composition approach for aerial-terrestrial edge-cloud networks. The networks, particularly leveraging UAVs as ABSs, offer unparalleled opportunities for addressing user mobility challenges and enabling ubiquitous service access. The problem was modeled as an optimization framework, addressing ABS non-determinism and user mobility with deep reinforcement learning algorithms to predict ABS locations and service requests. By integrating predictive algorithms and isomorphic matching techniques, the method enabled cost-effective, latency-aware resource allocation. The simulation results confirmed the proposed method's superiority over baseline methods in terms of service request admission, cost, and E2E latency. The mechanism maintained service continuity with minimal unsupported requests, showcasing its scalability and robustness which highlighted its potential to enhance service orchestration in futuristic networks. Future work includes enhancing predictive capabilities and exploring resource allocation in the quantum internet \cite{quantumComputing} as well as multi-objective optimization for energy efficiency and fairness.

% \vspace{-2pt}
\section*{Acknowledgment}
% \vspace{-2pt}
This research work is partially supported by the Business Finland 6Bridge 6Core project under Grant No. 8410/31/2022, the European Union’s Horizon Europe research and innovation programme under the 6G-SANDBOX project with Grant Agreement No. 101096328, and the Research Council of Finland 6G Flagship Programme under Grant No. 369116.

\vspace{-2pt}
\bibliographystyle{IEEEtran}
\bibliography{conf_short,IEEEabrv,main}

\end{document}